# Innovation in Education: Developing and Assessing Gamification in the University of the Philippines Open University Massive Open Online Courses

Cecille Moldez[1], Mari Anjeli Crisanto[1], Ma. Gian Rose Cerdeña[1], Diego S. Maranan[1], Roberto Figueroa[1]

[1]Faculty of Information and Communication Studies, UP Open University, Los Baños, Philippines

***Corresponding author. Email:** cecille.moldez@upou.edu.ph

## Abstract

*The University of the Philippines Open University has been at the forefront of providing Massive Open Online Courses to address knowledge and skill gaps, aiming to make education accessible and contributing to societal goals. Recognising challenges in student engagement and completion rates within Massive Open Online Courses, the authors conducted a study by incorporating gamification into one of the University of the Philippines Open University's Massive Open Online Courses to assess its impact on these aspects. Gamification involves integrating game elements to motivate and engage users. This study explored the incorporation of Moodle elements such as badges, leaderboards, and progress bars. Using Moodle analytics, the study also tracked student engagement, views, and posts throughout the course, offering valuable insights into the influence of gamification on user behaviour. Furthermore, the study delved into participant feedback gathered through post-evaluation surveys, providing a comprehensive understanding of their experiences with the gamified course design. With a 28.86% completion rate and positive participant reception, the study concluded that gamification can enhance learner motivation, participation, and overall satisfaction. This research contributes to the ongoing discourse on innovative educational methods, positioning gamification as a promising avenue for creating interactive and impactful online learning experiences in the Philippines and beyond.*

## 1. Introduction

### 1.1. Background of the Study

Even before the COVID-19 pandemic, educational providers felt the need for accessible learning platforms that did not require physical attendance. One such platform that has helped fulfil this need is Massive Open Online Courses (MOOCs). A MOOC is a complete course consisting of educational content, assessments, peer-to-peer tutoring, and limited academic tutoring (Jansen et al., 2017).

MOOCs are online platforms whose main characteristics are (a) being open, where enrolment is free to anyone who has access to the Internet, and the learning pace is dictated by its user; (b) participatory, where learners may interact with their fellow learners and instructor and participate in the various learning activities prepared; and (c) distributed, where knowledge sharing is encouraged to foster creative thinking among its participants. These three characteristics lead to MOOCs being offered worldwide, without limit to who can register and enrol in a course (Baturay, 2015).

In 2012, the University of the Philippines Open University (UPOU), drawing on over a decade of experience in distance e-learning, introduced MOOCs in response to identified gaps in the Philippines (Bandalaria & Figueroa, 2018). The initiative addressed several crucial issues, including mismatched knowledge and skills between college graduates and industry needs, leading to unemployment and underemployment. Additionally, the MOOC initiative targeted school leavers, aiming to provide them with training for economic engagement. UPOU has since continued to offer online courses on various topics based on its targeted stakeholders and relevance to the current situation, following the university's mission to provide greater access to quality education and supporting Republic Act 10650 - Open Distance Learning Law. Table 1 summarises UPOU MOOC offerings from 2013 to 2018.

**Table 1.** Summary of UPOU MOOC Offerings from 2013 to 2018

| Year | No. of Courses Offered | No. of Enrolled Students | No. of Completers |
|---|---|---|---|
| 2013 | 1 | 390 | n/a |
| 2014 | 1 | 859 | n/a |
| 2015 | 10 | 2,547 | 48 |
| 2016 | 7 | 857 | 110 |
| 2017 | 23 | 1,741 | 154 |
| 2018 | 38 | 2,251 | 441 |

*Note.* Source: Almodiel et al., 2020

As evident in the summary in Table 1, the completion rates of MOOCs are typically low. This is problematic given that the completion rate indicates MOOCs' success (Rai & Chunrao, 2016). From the 76 MOOCs offered by UPOU from 2015 to 2018, the completion rate reported was 10.18% (Almodiel et al., 2020). This rate did not exceed that of other international MOOC providers. A recent study indicates that the average completion rates for MOOCs range from seven to ten percent (Fu et al., 2021; Gütl et al., 2014 as cited in Duncan et al., 2022), and it is uncommon for completion rates to surpass 25% (Jordan, 2015 as cited in Duncan et al., 2022). Findings from MOOC-related research and local experts have identified poor course design, specifically the lack of collaborative elements, as one of the challenges among the six emerging MOOC providers within the countries of the Asia-Pacific Economic Cooperation (APEC): Thailand, the Philippines, Malaysia, Indonesia, Vietnam, and Mexico (Jung et al., 2020). Additionally, the UK's governmental report on MOOCs emphasises that one of the reasons for this is the poor engagement of weaker learners (Rai & Chunrao, 2016). There are greater distractions to students' attention in a virtual classroom compared to a traditional, in-person classroom setup (Poondej & Lerdpornkulrat, 2020), To improve learner engagement in an online class, course providers can incorporate different features of a learning management system (LMS), including gamification.

Gamification, defined as integrating game elements into non-game contexts, operates on five levels: interface designs, design patterns, design principles, models, and design methods (Khalil et al., 2018). Learning platforms commonly employ badges, leaderboards, progress bars, experiences, and level challenges (Khalil et al., 2018; Poondej & Lerdpornkulrat, 2020). These elements aim to enhance user experience and boost engagement toward specific goals in learning environments, thus encouraging learners to actively participate in course activities (Aparicio et al., 2019). According to Kyewski and Krämer (2018), motivation is a pivotal factor in learning, and gamified elements, such as badge rewards, prove effective in heightening learners' motivation.

This case study incorporated gamification in the instructional design of a MOOC on Artificial Intelligence for Quality Assurance in Education.

## 1.2. Statement of the Problem

In consideration of challenges in student engagement and completion rates within MOOCs, this study attempted to determine how gamification influences learner engagement, completion rates, and overall satisfaction. Specifically, the study sought to address the following issues:
i. To what extent does the incorporation of gamification elements in UPOU MOOCs influence learner engagement levels, as reflected in metrics such as interaction frequency, participation rates, and time spent on course activities?
ii. How does the integration of gamification affect student completion rates within UPOU MOOCs?
iii. What are the perceived benefits and challenges of incorporating gamification elements, such as badges, leaderboards, and progress bars, in shaping participants' learning experiences in UPOU MOOCs?

## 1.3. Objectives

This case study sought to develop and assess gamification elements incorporated within one MOOC, specifically focusing on determining the potential impact of gamification on student engagement and the course's completion rate.

The specific objectives of this study were as follows:
i. Introduce gamification elements, i.e., badges, activity completion, experience points, leaderboards, and levels, in the MOOC on Artificial Intelligence for Quality Assurance in Education.
ii. Utilise Moodle analytics to systematically track and describe student engagement, views, and posts throughout the course.
iii. Analyse the completion rate of the MOOC on Artificial Intelligence for Quality Assurance in Education.

## 1.4. Significance of the Study

This study holds significant implications for various stakeholders in the educational landscape, including students, faculty members, UPOU, and other institutions offering MOOCs. Understanding how gamification influences engagement and satisfaction can improve students' learning experiences and outcomes. Faculty members can benefit by tailoring their teaching approaches to create more engaging online courses, enhancing student engagement and learning outcomes. The university can use the findings to inform institutional strategies and policies related to online education, contributing to its reputation as a leader in innovative education delivery. Additionally, other universities offering MOOCs can leverage these insights to enhance their online learning platforms and better support their own students' educational journeys.

## 1.5. Scope and Limitations

This study implemented gamification elements in the Moodle LMS within the UPOU MOOC titled "Artificial Intelligence for Quality Assurance in Education," conducted from 1 to 31 August 2022. The participation of each enrolled learner was tracked using Moodle analytics. Students were classified as completers if they fulfilled the final requirements, including submitting the final document and conducting peer evaluations, on or before 11 September 2022.

Since this study was conducted before the year UPOU saw a sudden increase in MOOC enrolment (as described in the study by Cerdeña et al., 2024), the sample size limits the generalisability of the study's findings. Nonetheless, the findings provide useful insights on how gamification can be further integrated into future MOOC offerings.

## 2. Literature Review

### 2.1. Learner Engagement in MOOCs

Also characterised as 'student engagement' (Trowler, 2010), the notion of 'learner engagement' is described by Deng et al. (2020) to refer specifically to course engagement in MOOCs. Learner engagement has four essential dimensions: (a) behavioural, referring to the participation in and completion of a learning activity; (b) cognitive, pertaining to the self-regulated learning achieved by a MOOC learner; (c) emotional, the sense of belonging that a learner experiences during the course; and (d) social engagement, meaning the interactions either with other learners or the teacher/facilitator (Deng et al., 2020; Klassen et al, 2013; Fredricks et al., 2004). Given that MOOC completion rates tend to be low and dropout rates high (Khalil & Ebner, 2014; Jarnac et al. da Silva, 2023), there is a gap in learner engagement in MOOCs. Various studies have identified factors that can affect learner engagement in MOOCs, such as the technological tools used, peer interaction, timely feedback, connectivity, and digital skills (Sun et al., 2020; Alemayehu & Chen, 2023; Walji et al., 2016).

The LMS chosen as the MOOC platform is a critical factor, as argued, for instance, by Zanjani (2017) in a study evaluating the LMS tool Blackboard. The study identified aspects of user interface and user experience design that affected engagement, such as the presence of too many tools and links, and the lack of a notification procedure. The results highlighted the need for a student-centred LMS design to enhance learner engagement (Zanjani, 2017). As Alemayehu and Chen (2023) also note, a poor course design with a lack of student interaction will only lessen a learner's motivation to engage in the course further, much less complete it. However, given that engagement is "the willingness of people to be active through interaction with the content and the people in the course" (Walji et al., 2016), the LMS is only part of the equation. Learners must also be encouraged or motivated to interact with the course or their peers through various activities. While certain LMS features and digital tools used in MOOC platforms might enhance learner engagement, the converse is also true: High learner engagement might lead to "more enthusiasm in technological tools to learn in MOOCs" (Sun et al., 2020).

### 2.2. Gamification in MOOCs

Given current trends in completion and dropout rates in MOOCs, it is unsurprising that gamification has been applied and investigated as a strategy to enhance learner engagement. Indeed, not only is gamifiability considered one of the most important features in any LMS, but it is also regarded as one of the most important instructional models of today (Poondej & Lerdpornkulrat, 2020). Gamification is the application of game elements in contexts that would usually not be considered a game setting (Khalil et al., 2018; Domínguez et al., 2013) in order to encourage desired behaviours (Popp & Schuhbauer, 2023). It draws inspiration mainly from the gaming culture, including popular video games from the early 2000s (Sheldon, 2011; Deterding et al., 2011). The overall goal of gamification is to enhance autonomy (Saputro & Salam, 2024), intrinsic motivation, and student engagement in order to reduce drop-out rates (Khalil et al., 2018), as well as increase the student's learning persistence until the end of the course (Cheng, 2024).

In a seminal paper on the topic, Deterding et al. (2011) suggest that gamification design can be understood along five levels of abstraction: interface designs, design patterns, design principles, models, and design methods (Khalil et al., 2018). Papadimitriou's (2024) systematic review identified fifteen gamification strategies in the literature: challenge-based, immersive, goal-oriented, social, attention-grabbing, persuasive, competitive, story-driven, progressive, service-oriented, card-based, augmented reality-based, guided discovery-based, team-based, and theory-based.

Commonly used gamification elements in learning platforms include badges, leaderboards, progress bars, experience points, and level challenges (Popp & Schuhbauer, 2023; Al-Hafdi & Alhalafawy, 2024). These elements are incorporated into learning environments to create a better user experience and increase engagement to achieve a specific goal to influence learners to take an active role in course activities (Aparicio et al., 2019). Badges are forms of reward that take the form of digital images indicating that the learner has achieved some predetermined goal (Popp & Schuhbauer, 2023; Moskal et al., 2015). Gamified elements such as badges can effectively increase learners' intrinsic and extrinsic motivation (Kyewski &

Krämer, 2018; Popp and Schuhbauer, 2023), although the effects were not always observed in the studies listed in one recent systematic review (Popp & Schuhbauer, 2023). Leaderboards are ordered, ranked displays of how users perform in comparison to each other (Jia et al., 2017). Leaderboards appear to influence goal-setting and commitment (Landers et al., 2017; Popp & Schuhbauer, 2023). However, this is not the case if the leaderboards are relative, i.e., showing only a learner's relative rank compared to the person directly above or below them on a ranked list (Popp & Schuhbauer, 2023). Progress bars visually depict the extent to which a task has been accomplished and indicate how much more work is still needed (Mazarakis & Bräuer, 2023). Experience points are scores that monotonically increase toward some maximum possible value (Gehringer & Peddycord, 2013).

Recent systematic literature reviews agree on the potential of gamification in MOOCs in increasing user participation, such as in getting more tasks done, and improving forum participation, motivation, and achieving learning outcomes (Jarnac et al. da Silva; 2023; Saputro & Salam, 2024; Al-Hafdi & Alhalafawy, 2024). Several studies have also shown that a gamified context in MOOCs leads to a higher level of participation and enjoyment and is an overall significant contributing factor to the overall success of a MOOC (Aparicio et al., 2019; Cheng, 2021; Ortega-Arranz et al., 2019; Vaibhav & Gupta, 2014). Furthermore, comparative studies between non-gamified and gamified learning environments have found that with the latter, there is a decrease in learner dropouts and greater 'visible' presence from the learners; if gamified, the learning platform even manages to encourage learners to perform optional tasks (Cheng, 2021; Ortega-Arranz et al., 2019; Vaibhav & Gupta, 2014). Gamification can yield higher grades due to higher retention during classes and has even been regarded in some cases as essential to achieving learning outcomes (Pacatang & Pandi-Ruedas, 2024). In short, gamification in a MOOC context can reduce student dropout rates, improve learner satisfaction and user experience (Aparicio et al., 2019), and increase learner motivation. However, correctly implementing gamified elements can be difficult (Kyewski & Krämer, 2018; Mazarakis & Bräuer, 2023). Presentation is a particularly important factor for successful retention of users, as a good presentation of learning materials increases users' interest and engagement in the course, leading to a higher success rate during assessment (Vaibhav & Gupta, 2014). It has also been recorded that users tend to find assessment "boring and tedious" in a non-gamified learning environment (Vaibhav & Gupta, 2014). Notably, users who do not aim to complete a MOOC in the first place will neither be engaged nor motivated by rewards presented in a gamified environment (Jarnac de Freitas & Mira da Silva, 2023). In addition, gamified learning experiences can lead to unwanted outcomes, such as increased competition and difficulty evaluating task completion (Hamari et al., 2014; Domínguez et al., 2013). Finally, it is worth noting that the experience of gamified courses can be hampered by factors external to the gamification design or learner profiles, such as slow Internet connection, or difficulties in the deployment of digital tools if implemented in a traditional learning environment (Medico et al., 2023; Sipin & Tan, 2023; Pacatang & Pandi-Ruedas, 2024; Esparrago-Kalidas et al., 2024). The relationship between personal traits and preferences and the choice of gamified strategies for the most impact still does not appear well understood (Pulist & Sharma, 2024). Perhaps it is for all these reasons that gamified MOOCs with a dedication to increasing completion rates through personalised learning paths are still few and far between (Jarnac et al. da Silva, 2023; Antonaci et al., 2017).

## 3. Methodology

This section details the creation of the gamified e-learning component using Moodle, offering a comprehensive explanation of the developmental process. Key phases encompassed planning, implementation, and evaluation, highlighting critical aspects of the overall development journey.

### 3.1. Participants

This study was conducted within the UPOU MOOC on Artificial Intelligence for Quality Assurance in Education, which was made available from 1 to 31 August 1 to 2022. During this period, 201 students participated in this gamified MOOC.

### 3.2. Platform

UPOU's MODeL is the primary LMS for UPOU's MOOC offerings. Given that Moodle is a suitable platform to implement gamified e-learning (Poondej & Lerdpornkulrat, 2020), as long as gamification elements are applied carefully and intentionally (Reif et al., 2024), it was possible to incorporate features conducive to a gamified version of an earlier version of the course site. Before launching the MOOC, the authors integrated game-like elements and categories to enhance student engagement. As shown in Figure 1, the UPOU Model was configured specifically for this study.

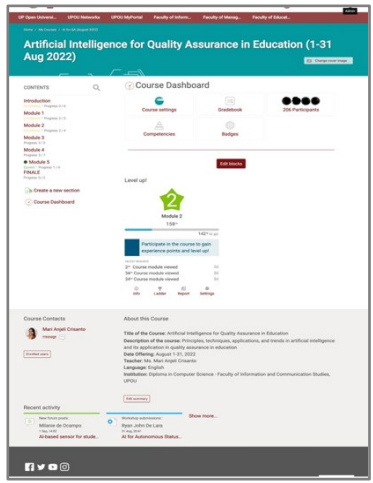

**Figure 1.** Course Dashboard　　　　　　**Figure 2.** Course Report Page

### 3.3. Planning Phase

#### 3.3.1. Course Planning

The MOOC included an introductory module, five topic modules, and a finale. The introductory module, accessible before the course began, allowed learners to familiarise with gamification and course rules. It featured a welcome video explaining gamification and key terms. To enhance the game-like atmosphere, learners were called 'players' and facilitators 'game masters.'

#### 3.1.2. Gamification Planning

To encourage the students' active participation, the authors identified the following game elements used in the MOOC learning context: experience points (XP), activity completion, levels, badges, and leaderboards.

##### 3.1.2.1. Experience Points (XP)

The authors pinpointed key game elements integrated into the MOOC learning environment to foster active student engagement, including XP, activity completion, levels, badges, and leaderboards.

##### 3.1.2.2. Activity Completion

A notable aspect of Moodle is its activity completion feature, which can be beneficial in guiding learners through their tasks. The system automatically marked a tick box next to each activity upon learners' fulfilling specific requirements, offering clear guidance to the learners on their progress.

*3.1.2.3.   Levels*

The MOOC was designed like a game, with five modules and a final assessment, for a total of six levels. Students earned points based on their performance in each module's tasks. Completing more activities in a module meant earning more points, similar to advancing in a game. Course facilitators, called 'game masters,' could monitor students' progress using the course report shown in Figure 2.

*3.1.2.4.   Badges*

Learners earned badges to recognise every significant accomplishment. These badges were automatically given when learners completed tasks such as participating in all discussion forums, watching every video lecture, reading all course materials, and submitting assessments. There were 11 badges designed for the course, including badges for completing levels and a welcome badge for all participants.

*3.1.2.5.   Leaderboards*

This feature visually displays each learner's rank within the class. While students could view their ranks, they could not see those of their classmates. The ranking system was designed to maintain anonymity, safeguarding the identities of other learners in a MOOC offering of this nature. An example of this is shown in Figure 3.

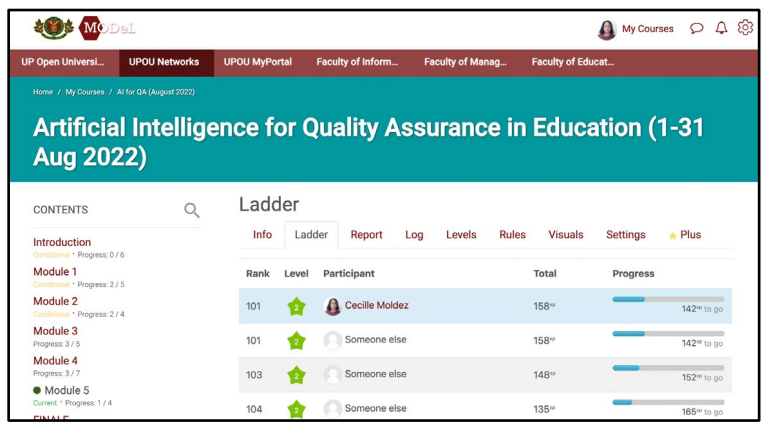

**Figure 3.** Level Up! Page

**3.2.   Implementation Phase**

UPOU has consistently utilised Moodle as a principal LMS for most courses. Leveraging Moodle's inherent capabilities for gamification, supplemented by the 'Level-up!' plug-in (https://www.moodle.org/plugins/block_xp), facilitated the seamless implementation of gamification plans. The authors strategically selected and integrated fundamental Moodle activities to handle MOOC content on the site. The table below provides a detailed breakdown of each utilised Moodle feature, incorporating various game elements.

**Table 2.** Moodle Features

| Features in Moodle Setup | |
|---|---|
| Activity Completion | The course site was configured to recognise fulfilled conditions, automatically marking activities as complete (indicated by a tick box) once all specified criteria were met. |
| Badges | The badges were designed to be automatically awarded to learners who meet assigned criteria, such as completing an activity or reaching a significant milestone. |
| Forums | Learners actively engaged with each other in discussion forums, providing a platform for social interaction and collaboration. The course facilitator initiated these forums with questions, fostering an environment for students to connect with their peers. |

| **Features in Moodle Setup** | |
|---|---|
| URL Links | Resource materials, including video lectures, readings, and downloadable content, were thoughtfully organised within each module. The system diligently tracked every instance of accessing these materials, attributing points to students accordingly. |
| Assessment | The Moodle Workshop feature is a valuable tool for fostering hands-on and interactive learning among students. Students submit their course assignments, and their peers and facilitator review them to provide ratings based on predefined criteria. |
| Level Up! Block Plug-in | The Moodle plug-in assigns XP to learners based on specific course activities, including forum participation, accessing reading materials, viewing video lectures, engaging in surveys, and completing course assessments. Additionally, it helps students monitor their own progress. |

The MOOC comprised five modules, the accessibility of which were scheduled weekly. The course managers utilised these modules as the foundation for level settings, including an additional final level. In total, this gamified course featured six levels. Notably, the MOOC participants constitute diverse age groups; this likely influenced their perceptions and interactions with the gamification elements differently.

### 3.3. Evaluation Phase

The gamification of the UPOU MOOC on Artificial Intelligence for Quality Assurance in Education made learning more fun and personalised for the students. Using tools like Moodle analytics and game elements ensured that students felt a sense of accomplishment when they participated.

An in-depth analysis of the course site views and posts from Moodle reports, which demonstrated the students' engagement, was subsequently conducted. The correlation between student engagement from their respective XP and completion rate characterised by the badges was especially emphasised. Furthermore, this study focused on specific gamification elements such as badges, leaderboards, and XP, which might not encompass all possible gamification strategies such as the inclusion of more immersive and interactive elements like virtual reality (VR) and interactive scenarios. The final requirements included a peer evaluation and course survey. The survey aimed to obtain an overall assessment of the MOOC, thus only two items specifically addressed the course's learning design and gamification aspects.

## 4. Findings and Discussion

Out of the 201 enrolled students, a maximum number of 145 students actively participated on the course site. The study utilised Moodle analytics to analyse their interactions and engagement behaviours. However, only 85 out of the 145 participants responded to the post-evaluation survey that was deployed to seek feedback on the course design and gamified course experience.

### 4.1. Student Engagement based on Moodle Analytics

To monitor student participation, the researchers documented the views for each activity in the MOOC and tracked the number of total active users using reports from Moodle. It was essential to analyse the trend in student engagement from the initial week to the fourth week of the course. Table 3 summarises the total views for each activity per week and the overall number of users accessing these activities throughout the specified period.

**Table 3.** Total Views and Active Users Per Week

| MOOC Activity | Week 1* | | Week 2* | | Week 3* | | Week 4* | | Week 5* | |
|---|---|---|---|---|---|---|---|---|---|---|
| | Views | Users | Views | Users | Views | Users | Views | Users | Views | Users |
| Level 1 Checkpoint | 403 | 114 | 468 | 132 | 492 | 139 | 502 | 140 | 514 | 144 |
| Self Introduction | 1,174 | 120 | 2,019 | 135 | 2,268 | 142 | 2,484 | 142 | 2,948 | 145 |
| Module 1 Video Link | 204 | 113 | 245 | 129 | 274 | 136 | 284 | 138 | 292 | 141 |
| Discussion Forum 1 | 2,797 | 109 | 3,926 | 129 | 4,691 | 135 | 4,933 | 137 | 5,566 | 140 |
| Module 1 Study Guide | 200 | 108 | 263 | 124 | 288 | 131 | 279 | 134 | 306 | 135 |
| Level 2 Checkpoint | | | 189 | 83 | 253 | 108 | 271 | 113 | 288 | 117 |
| Module 2 Video Link | | | 154 | 77 | 210 | 105 | 224 | 110 | 234 | 116 |
| Discussion Forum 2 | | | 1,885 | 74 | 2,975 | 103 | 3,383 | 111 | 3,728 | 115 |
| Module 2 Study Guide | | | 113 | 65 | 169 | 96 | 181 | 103 | 195 | 108 |
| Level 3 Checkpoint | | | | | 139 | 62 | 199 | 86 | 225 | 95 |
| Module 3 Video part 1 | | | | | 107 | 65 | 159 | 87 | 179 | 94 |
| Module 3 Video part 2 | | | | | 96 | 60 | 142 | 82 | 157 | 88 |
| Discussion Forum 3 | | | | | 1,582 | 58 | 2,394 | 82 | 2,811 | 90 |
| Module 3 Study Guide | | | | | 82 | 55 | 132 | 80 | 148 | 89 |
| Level 4 Checkpoint | | | | | | | 173 | 70 | 225 | 85 |
| Module 4 Video Link | | | | | | | 103 | 68 | 132 | 86 |
| Reading Material 1 | | | | | | | 103 | 66 | 133 | 82 |
| Reading Material 2 | | | | | | | 95 | 63 | 121 | 80 |
| Discussion Forum 4 | | | | | | | 1,397 | 61 | 2,141 | 78 |
| Module 4 Study Guide | | | | | | | 85 | 58 | 116 | 75 |
| MOOCs Survey | | | | | | | 116 | 62 | 156 | 75 |
| Level 5 Checkpoint | | | | | | | | | 136 | 65 |
| Module 5 Video Link | | | | | | | | | 99 | 67 |
| Discussion Forum 5 | | | | | | | | | 1,245 | 68 |
| Module 5 Study Guide | | | | | | | | | 96 | 66 |
| Final Assignment | | | | | | | | | 107 | 41 |

It is crucial to note that each module corresponds to each week of the MOOC offering. The data was captured and analysed from the reports generated by the Moodle platform. Analysing the data presented in Table 3 shows key trends in student involvement. Out of the 201 enrolled students, a maximum of 145 actively engaged with the course site during the fifth week, with the highest activity recorded in the Self

Introduction activity. The predominant role of discussion forum activities in driving engagement is noteworthy, which is evident in their substantial views. By the fifth week, the Self Introduction activity had 2,948 views, Discussion Forum (DF) 1 amassed 5,566 views, DF2 received 3,728 views, DF3 had 2,811 views, DF4 recorded 2,141 views, and DF5 had 1,245 views. This underscores the significance of discussion forums as pivotal platforms for fostering student interaction and participation. Figures 4 and 5 show the trend of the views and students' posts for the entire MOOC offering.

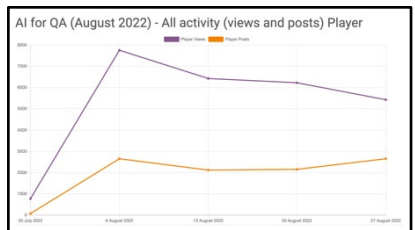

**Figure 4.** Student engagement

| Period ending (Week) | Player Views | Player Posts | Logs |
|---|---|---|---|
| 27 August 2022 | 5426 | 2645 | Course Logs |
| 20 August 2022 | 6221 | 2148 | Course Logs |
| 13 August 2022 | 6415 | 2111 | Course Logs |
| 6 August 2022 | 7755 | 2650 | Course Logs |
| 30 July 2022 | 768 | 72 | Course Logs |

**Figure 5.** Weekly views and posts as recorded by Moodle based on views and posts

### 4.2. Student Engagement Inferred from MOOC Badges Received

Issuing badges in the course was a reward mechanism for learners who achieved set criteria and milestones. Table 4 outlines the variety of badges conferred and the corresponding number of recipients for each badge as reported by Moodle. The distribution of badges not only recognised individual accomplishments but also contributed to fostering a sense of achievement and motivation among participants. The badges also indicated that certain activities were successfully completed by the participants.

**Table 4.** Summary of Badge Recipients

| Badge Name | Description | Total Number of Recipients |
|---|---|---|
| Newbie Badge | Awarded to a student who successfully enrolled in this MOOC. | 201 |
| Digital Analyzer | Awarded to a student who watched all the video lectures. | 83 |
| Digital Bookworm | Awarded to a student who read all the required reading materials. | 94 |
| Digital Comm. | Awarded to a student who participated in all Discussion Forums. | 80 |
| Level 1 Completer | Awarded to a student who completed all the activities in Level 1. | 117 |
| Level 2 Completer | Awarded to a student who completed all the activities in Level 2. | 100 |
| Level 3 Completer | Awarded to a student who completed all the activities in Level 3. | 86 |
| Level 4 Completer | Awarded to a student who completed all the activities in Level 4. | 82 |
| Level 5 Completer | Awarded to a student who completed all the activities in Level 5. | 76 |
| FINALE Badge | Awarded to a student who is eligible to take the Finale Level. | 76 |
| AI for QA Badge | Awarded to a student who successfully completed this MOOC. | 58 |

In addition to the introductory 'Newbie' badge, the most frequently earned badge was the 'Level 1 Completer', which had 117 recipients. Noticeably, the number of badges declined towards the end of the course. The data in Table 4 highlights the significance of gamification elements, such as badges, in monitoring activity completion throughout the course.

Furthermore, it is imperative to ascertain the connection between accomplishing tasks and engaging students to cultivate productive learning experiences. Considering all 201 students enrolled in this course, the study provided an overview of activity completion and student engagement through the accumulation of badges and XP.

**Table 5.** Badge Distribution

| Accumulated badges | No. of students |
|---|---|
| 11 | 67 |
| 10 | 8 |
| 9 | 1 |
| 8 | 3 |
| 7 | 1 |
| 6 | 2 |
| 5 | 4 |
| 4 | 5 |
| 3 | 12 |
| 2 | 14 |
| 1 | 84 |

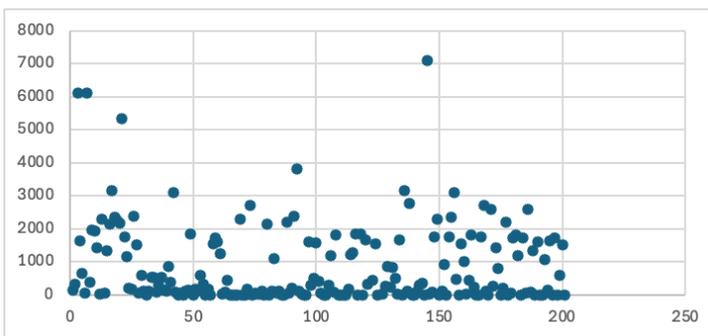

**Figure 6.** XP Distribution

Table 5 and Figure 6 show the distribution of badges and XP, respectively. The first column indicates the number of badges earned by each student, ranging from 1 to 11, reflecting varying levels of activity completion. Figure 6 shows the XP accumulated by each student (ranging from 0 to 7,114), indicating student engagement and activity. Analysing the correlation between badges earned and XP can provide valuable insights into the relationship between task completion and student engagement. The correlation coefficient (Pearson's r) can be calculated to determine the degree and direction of the association between activity completion and student involvement.

The activity completion and student engagement were strongly positively correlated, $r(199) = .77$, $p < .01$. This study indicates a significant and positive connection between the number of badges obtained by students and their corresponding XP. This exciting correlation demonstrates that students who actively completed more tasks – as indicated by their badge count – generally exhibited higher levels of engagement.

### 4.3. Completion of MOOC Modules and Activities

Moodle can generate the actual number of badges given out for each module. The average number of users per module can be computed from the number of users in the fifth week, categorised by the modules to which they belong, as seen in Table 5. The summary of the results is presented in Table 6. From the data, the estimated completion rate for each week can be computed using the formula below.

*(Actual number of badge completion recipients per module/Average number of users who accessed the module) * 100 = Estimated completion rate per module*

**Table 6.** Estimated Completion Rate Per Module

| Module Number | Average number of badge completion recipients per module | Average number of users who accessed the module | Estimated completion rate per module |
|---|---|---|---|
| Module 1 | 114 | 141 | 81% |
| Module 2 | 95 | 114 | 83% |
| Module 3 | 79 | 91 | 87% |
| Module 4 | 70 | 80 | 87% |
| Module 5 | 58 | 67 | 87% |

The completion rates for each module were commendable. Specifically, Module 1 achieved an 81% completion rate, Module 2 an 83% completion rate, Module 3 an 87% completion rate, Module 4 an 87% completion rate, and Module 5 also an 87% completion rate. These figures underscore the overall success and engagement of learners across the different modules of the course.

Despite a decline in the number of learners receiving badges for completing each module, it is noteworthy that the completion rate, calculated based on the number of learners who accessed the module, exhibited an upward trend as the course advanced. This positive trajectory suggests that the learners, while fewer in number, were increasingly successful in meeting the module requirements, emphasising their growing commitment to engagement and accomplishment as the MOOC unfolded. Additionally, this trend may indicate a higher level of dedication and involvement among those who remained active in the later stages of the course.

The overall completion rate for the MOOC was determined by calculating the percentage of the number of students who received a 'Completion Certificate' against the total number of officially enrolled students. To earn this certificate, learners had to submit the final requirement and evaluate their peers' submissions. Out of 201 enrolled students, 58 met all criteria, yielding a completion rate of 28.86%. Although 66 students submitted the final requirements, nine failed to complete the peer review process and were thus not considered among the completers.

### 4.4. Gamification Feedback and Suggestions

The feedback gathered from participants provided valuable insights into their experiences with the gamified MOOC. These comments reflect diverse perspectives on the course design, user interface, and overall learning environment.

While some lauded the innovative approach, praising how gamification ensured consistent participation throughout the course, others found themselves grappling with distractions brought about by the game mode. One participant, in particular, expressed their discontent, noting that the gamified elements felt more obligatory than engaging, making it challenging to fully immerse themselves in the learning process. This suggests a need for careful balancing of game elements to ensure they support rather than hinder learning.

Amidst the mixed reviews, a common thread emerged: the desire for more stimulating teaching methods. Some participants acknowledged the quality of the course content but expressed hope for a more engaging and interactive learning environment in the future. While gamification has its merits, on its own it may not be sufficient to keep all students fully engaged. This highlights the importance of diversifying teaching methods and integrating more immersive and interactive technologies. A selection of the participants' comments and recommendations are listed below.

**Consistent Participation and Engagement:**
- "This is my first time trying gamification, and I commend how it ensured that participation remains consistent throughout the course."
- "I love that it was gamified so I was kind of forced to really study and engage with other classmates."

**Interactive Learning and Engagement:**
- "The course is interactive as it is designed to have a point system. I learned a lot from the instructor aided with the materials."
- "It has a good user interface and user experience. I was able to learn and understand the lesson because it has the tools needed."

**Positive Learning Experience and Application:**
- "Engaging, interesting, very informative. The course, together with the instructor, is very good."

- "Gamification of this course is quite an experience and I may be able to apply it in the delivery of courses assigned to me."

**Feedback on Gamification:**
- "The game mode is distracting."
- "I think the gaining of points was a little difficult. It required me to comment on other people's work just because 'I need to,' not because I feel particularly engaged."

**Desire for More Stimulating Teaching Methods:**
- "Please do not take this the wrong way, but the method of teaching here is quite... unstimulating. I have other online learnings from the company and those are in forms of VR like games, interactive videos, interactive scenarios. The topics here are really good but I hope it will be more fun in the future."

### 4.5. Recommendations

In the future, the course design for the gamified MOOC will explore integrating additional interactive learning strategies. This may include incorporating suggested elements such as VR games, interactive videos, and scenarios that can enhance engagement and provide a more immersive learning experience.

Furthermore, universities can collaborate to develop best practices for gamified learning environments, sharing insights and resources to improve the quality of online education. Educational institutions can use the findings to inform policy development for online course delivery, emphasising the importance of innovative and interactive learning methods. Furthermore, institutions can invest in technological infrastructure to support advanced gamification elements like VR and augmented reality (AR), creating immersive and engaging learning experiences.

## 5. Conclusion

In the culmination of this study, a critical examination of student engagement within the gamified MOOC reveals essential insights crucial to understanding the efficacy of the instructional design. Each module, aligned with the corresponding week of the MOOC offering, is a pivotal component in the structured learning experience. The data, captured and analysed through Moodle's reporting mechanisms, serves as a lens to scrutinise the trends in student participation, offering a nuanced view of their interactions with the course content. The module-specific completion rates, indicating success at each stage, collectively contributed to an overall completion rate of 28.86%, which stands out as notably positive compared to other MOOCs concurrently offered by the UPOU. The impact of gamification in fostering learner engagement is evident, as participants were motivated to partake in the various course activities. Further studies on integrating gamification into MOOCs with more than a thousand enrollees would be helpful in verifying the findings presented in this study. MOOCs with more participants could benefit from seeing how incorporating elements such as XP, activity completion, levels, badges, and leaderboards would contribute to MOOC completion at a larger scale. Additionally, future research should explore age-appropriate gamification in MOOCs, assess its long-term impact on learner engagement and completion rates, compare it with other strategies, investigate cultural differences, and explore integrating advanced gamification elements like VR and AR to enhance the learning experience.

**Funding:** This study was funded by the 2022 UPOU API entitled "Offering and Evaluation of DCS-initiated MOOCS".

**Acknowledgement:** The author would like to acknowledge the contributions of the journal advisors, chairpersons, editorial board members and the respective international offices for their continuous support. limited to grant providers and/or selected individuals whose work made a significant contribution to article presented.